\documentclass[preprint,aps,eqsecnum]{revtex4} 
\usepackage{bm,amsfonts,amsmath}

\begin{document}


\def\eqref#1{(\ref{#1})}
\def\eqrefs#1#2{(\ref{#1}) and~(\ref{#2})}
\def\eqsref#1#2{(\ref{#1}) to~(\ref{#2})}

\def\Eqref#1{Eq.~(\ref{#1})}
\def\Eqrefs#1#2{Eqs.~(\ref{#1}) and~(\ref{#2})}
\def\Eqsref#1#2{Eqs.~(\ref{#1}) to~(\ref{#2})}
\def\Eqpartref#1#2{Eq.~(\ref{#1}{#2})}

\def\Ref#1{Ref.~\cite{#1}}
\def\Refs#1{Refs.~\cite{#1}}

\def\proclaim#1{\medbreak
\noindent{\it {#1}}}
\def\Proclaim#1#2{\medbreak
\noindent{\bf {#1}}{\it {#2}}\par\medbreak}

\def\EQ #1\doneEQ{\begin{equation} #1 
\end{equation}}
\def\EQs #1\doneEQs{\begin{eqnarray} #1 
\end{eqnarray}}

\def\ontop#1#2{
\setbox2=\hbox{{$#2$}} \setbox1=\hbox{{$\scriptscriptstyle #1$}} 
\dimen1=0.5\wd2 
\advance\dimen1 by 0.5\wd1 
\dimen2=1.4\ht2
\ifdim\wd1>\wd2 \raise\dimen2\box1 \kern-\dimen1 \hbox to\dimen1{\box2\hfill}
\else \box2\kern-\dimen1 \raise\dimen2 \hbox to\dimen1{\box1\hfill} \fi }

\def\eqtext#1{\hbox{\rm{#1}}}

\def\mixedindices#1#2{{\mathstrut}^{#1}_{#2}}
\def\downindex#1{{\mathstrut}^{\mathstrut}_{#1}}
\def\upindex#1{{\mathstrut}_{\mathstrut}^{#1}}
\def\downupindices#1#2{\downindex{#1}\upindex{#2}}
\def\updownindices#1#2{\upindex{#1}\downindex{#2}}

\def\id#1#2{\delta\downupindices{{\internal #1}}{{\internal #2}}}
\def\cross#1#2{\epsilon\,\updownindices{#1}{#2}}
\def\vol#1{\epsilon\,\downindex{#1}}
\def\invvol#1{\epsilon\,\upindex{#1}}
\def\cross#1#2{\epsilon\,\downupindices{#1}{#2}}

\def\flat#1{\eta\downindex{#1}}
\def\invflat#1{\eta\upindex{#1}}

\def\sodder#1#2{\sigma\updownindices{#1}{#2}}
\def\invsodder#1#2{\sigma\downupindices{#1}{#2}}

\def\der#1{\partial\downindex{#1}}
\def\coder#1{\partial\upindex{#1}}

\def\ELop#1#2#3#4{{{\rm E}_h}({#4})\downupindices{#1}{#2{\internal #3}}}

\def\Lie#1{{\cal L}_{#1}}
\def\LieE#1#2#3{(\Lie{\delta}E)_{#3}\mixedindices{#2}{#1}}


\def\h#1#2#3{h\downupindices{#1}{#2{\internal #3}}}
\def\w#1#2#3{\omega\downupindices{#1}{#2{\internal #3}}}
\def\R#1#2#3{R\downupindices{#1}{#2{\internal #3}}}
\def\tw#1#2#3{{\tilde\omega}\downupindices{#1}{#2{\internal #3}}}

\def\W#1#2#3{\Omega\downupindices{#1}{#2{\internal #3}}}
\def\powW#1#2#3#4{\Omega^{#4}\downupindices{#1}{#2{\internal #3}}}
\def\newW#1#2#3{\Omega'\downupindices{#1}{#2{\internal #3}}}

\def\curlh#1#2#3#4{\Gamma\downupindices{#1}{#2}\downupindices{#3}{\internal #4}}

\def\symmh#1#2#3{\gamma\downupindices{#1}{#2{\internal #3}}}
\def\newsymmh#1#2#3{\gamma'\downupindices{#1}{#2{\internal #3}}}

\def\vect#1#2#3{\xi\downupindices{#1}{#2{\internal #3}}}
\def\spin#1#2#3{\chi\downupindices{#1}{#2{\internal #3}}}
\def\vectsub#1#2#3#4{\xi_{#1}\downupindices{#2}{#3{\internal #4}}}
\def\spinsub#1#2#3#4{\chi_{#1}\downupindices{#2}{#3{\internal #4}}}
\def\tspin#1#2#3{{\tilde\chi}\downupindices{#1}{#2{\internal #3}}}
\def\tspinsub#1#2#3#4{{\tilde\chi}_{#1}\downupindices{#2}{#3{\internal #4}}}

\def\newh#1#2#3#4{\ontop{#1}{h'}\downupindices{#2}{#3{\internal #4}}}
\def\newvect#1#2#3#4{\ontop{#1}{\xi'}\downupindices{#2}{#3{\internal #4}}}
\def\newspin#1#2#3#4{\ontop{#1}{\chi'}\downupindices{#2}{#3{\internal #4}}}

\def\varh#1#2#3#4#5{\ontop{(#2)}{\delta_{#1}}\h{#3}{#4}{#5}}
\def\varw#1#2#3#4#5{\ontop{(#2)}{\delta_{#1}}\w{#3}{#4}{#5}}
\def\varwtop#1#2#3#4#5#6{\ontop{(#2)}{\delta_{#1}}\wtop{#3}{#4}{#5}{#6}}

\def\var#1#2{\ontop{(#1)}{\delta_{#2}}}
\def\varcomm#1#2#3{\ontop{(#1)}{[\delta_{#2},\delta_{#3}]}}

\def\Rtop#1#2#3#4{\ontop{(#1)}{R}\downupindices{#2}{#3{\internal #4}}}
\def\wtop#1#2#3#4{\ontop{(#1)}{\omega}\downupindices{#2}{#3{\internal #4}}}

\def\Etop#1#2#3#4{\ontop{(#1)}{E}\downupindices{#2}{#3{\internal #4}}}

\def\Ltop#1{\ontop{(#1)}{L}}

\def\E#1#2#3{E\downupindices{#1}{#2{\internal #3}}}
\def\coE#1#2#3{E\updownindices{#1}{#2{\internal #3}}}
\def\newE#1#2#3{E'\downupindices{#1}{#2{\internal #3}}}

\def\T#1{T\downindex{#1}}

\def\onshell{|_{\ontop{\scriptscriptstyle (1)}{\scriptstyle E}=0}}

\def\internal{}

\def\a#1#2{a\updownindices{#1}{#2}}
\def\b#1#2{b\updownindices{#1}{#2}}
\def\c#1#2{c\updownindices{#1}{#2}}

\def\alg{{\mathcal A}}
\def\Dalg{{\mathcal A}_{(1)}}
\def\coDalg{{\mathcal A}'_{(1)}}
\def\PDalg{{\mathcal P}_{\Dalg}}
\def\PcoDalg{{\mathcal P}_{\coDalg}}
\def\PTDalg{\PDalg^{\rm T}}
\def\P#1#2{{\mathcal P}\mixedindices{#1}{#2}}
\def\PT#1#2{{\mathcal P}^{\rm T}\mixedindices{#1}{#2}}


\def\fflat#1{\bar\eta\downindex{#1}}
\def\finvflat#1{\bar\eta\upindex{#1}}
\def\fvol#1{\bar\epsilon\,\downindex{#1}}
\def\finvvol#1{\bar\epsilon\,\upindex{#1}}
\def\y#1#2{y\downupindices{#1}{#2}}

\def\fh#1#2#3{\bar h\downupindices{#1}{#2{\internal #3}}}
\def\fw#1#2#3{\bar \omega\downupindices{#1}{#2{\internal #3}}}
\def\fsymmh#1#2#3{\bar\gamma\downupindices{#1}{#2{\internal #3}}}

\def\fA#1#2{\bar A\downupindices{#1}{{\internal #2}}}
\def\fF#1#2#3{\bar F\downupindices{#1}{#2{\internal #3}}}
\def\fphi#1{\bar\phi\upindex{{\internal #1}}}
\def\fH#1#2#3{\bar H\downupindices{#1}{#2{\internal #3}}}

\def\fE#1#2#3{\bar E\downupindices{#1}{#2{\internal #3}}}
\def\fEtop#1#2#3{\ontop{(#1)}{\bar E}\downupindices{#2}{{\internal #3}}}
\def\fLtop#1{\ontop{(#1)}{\bar L}}

\def\fvect#1#2#3{\bar\xi\downupindices{#1}{#2{\internal #3}}}
\def\fspin#1#2#3{\bar\chi\downupindices{#1}{#2{\internal #3}}}

\def\fder#1{\bar\partial\downindex{#1}}


\def\Rnum{{\mathbb R}}

\def\unit#1{\openone\upindex{#1}}

\def\grav/{spin-two}
\def\ie/{i.e.}


\title{ Parity violating spin-two gauge theories }

\author{ Stephen C. Anco }
\affiliation{
Department of Mathematics, Brock University, St. Catharines, ON Canada L2S 3A1}
\email{sanco@brocku.ca}
\date{\today}

\begin{abstract}
Nonlinear covariant parity-violating deformations of 
free spin-two gauge theory are studied in $n\ge 3$ spacetime dimensions, 
using a linearized frame and spin-connection formalism, 
for a set of massless \grav/ fields. 
It is shown that the only such deformations 
yielding field equations with a second order quasilinear form 
are the novel algebra-valued types in $n=3$ and $n=5$ dimensions
already found in some recent related work
concentrating on lowest order deformations. 
The complete form of the deformation to all orders in $n=5$ dimensions
is worked out here and some features of 
the resulting new algebra-valued spin-two gauge theory are discussed. 
In particular, the internal algebra underlying this theory 
on $5$-dimensional Minkowski space 
is shown to cause the energy for the spin-two fields to be of indefinite sign.
Finally, 
a Kaluza-Klein reduction to $n=4$ dimensions is derived, 
giving a parity-violating nonlinear gauge theory of a
coupled set of \grav/, spin-one, and spin-zero massless fields. 
\end{abstract}

\maketitle

\section{ Introduction }

It is widely believed that the only allowed type of gauge symmetry
for a nonlinear massless \grav/ field is a diffeomorphism symmetry,
corresponding to a gravitational self-interaction of the field. 
Recent work \cite{Annalspaper,Henneaux1}
has addressed this question more generally 
for a set of any number of coupled massless \grav/ fields 
by the approach of deformations of linear abelian \grav/ gauge theory. 
A deformation, here, means adding linear and higher power terms
to the abelian \grav/ gauge symmetry
while also adding quadratic and higher power terms 
to the linear \grav/ field equation,
such that a gauge invariant action principle exists
which is not equivalent to the undeformed linear theory 
by nonlinear field redefinitions. 
The condition of gauge invariance has various formulations
\cite{AMSpaper,Henneaux2,brst}
that yield determining equations to solve for the allowed form of 
the deformation terms added order by order in powers of the fields. 
This approach, in contrast to earlier efforts in the literature,
makes no assumptions on the possible structure for 
the commutator algebra of the deformed gauge symmetries
and takes advantage of the necessary requirement that these gauge symmetries
should be realized by a nonlinear theory given by 
a deformation of the Lagrangian of the linear theory. 

The results of the deformation analysis 
in \Ref{Annalspaper}
show that if all deformation terms are required to involve 
no more derivatives of the \grav/ field and gauge symmetry parameter
than appear in the free theory, 
then in four dimensions 
the unique possible deformation for a set of one or more \grav/ fields
is given by an algebra-valued generalization of Einstein gravity theory 
\cite{Cutler-Wald,Wald}
based on a commutative, associative, invariant-normed algebra. 
As such a set of fields is mathematically equivalent to 
a single algebra-valued \grav/ field \cite{multigravpaper},
the only type of gauge symmetry indeed allowed is 
a diffeomorphism symmetry (in an algebra-valued setting). 
An extension of this result to all higher spacetime dimensions is
given in \Ref{Henneaux1}
by using a BRST cohomological formulation \cite{Henneaux2,brst}
of the deformation determining equations. 

Very interestingly, in \Ref{exoticth}
a deformation different than the Einstein field equation 
and diffeomorphism gauge symmetry 
for a single \grav/ field in three-dimensional Minkowski space 
is constructed 
by deforming the abelian gauge symmetry by 
a linear term that contains first derivatives of both 
the \grav/ field and the gauge symmetry parameter. 
Gauge invariance is maintained at lowest order 
by also deforming the free Lagrangian by 
a term that is cubic in first derivatives of the \grav/ field. 
The resulting quadratic terms in the field equation 
contain one more derivative than in the free theory
but are still second order in derivatives. 
These deformation terms, moreover, 
also involve the Minkowski volume tensor 
and hence possess the interesting feature of being parity non-invariant. 
Of course, in three dimensions
there are no local dynamical degrees of freedom for a free \grav/ field. 
Indeed the full deformation to all orders is found to be related 
through field redefinitions to certain 
topological three-dimensional gravity theories \cite{topologicalgravity}
in a scaling limit in which the gravitational interaction is turned off. 
Intriguingly, 
parity non-invariant deformation terms of a similar form to 
the three-dimensional ones 
also exist at lowest order in five dimensions
for an algebra-valued \grav/ field
using an anticommutative algebra, 
but it was left open in \Ref{exoticth}
whether a full deformation actually exists to all orders. 
The resulting nonlinear \grav/ gauge theory, if it were to exist,
would be of obvious potential physical and mathematical interest 
to investigate. 
It would lead, for instance, to an exotic four-dimensional gauge theory
of a nonlinearly coupled set of \grav/ and spin-one fields
obtained via a Kaluza-Klein reduction of the five-dimensional theory. 

The main purpose of this paper is to show that, in fact,
the novel first order deformation in five dimensions 
explored in \Ref{exoticth}
exists to all orders only if the anticommutative algebra is
nilpotent (and invariant-normed), 
due to an integrability condition that arises in solving 
the deformation determining equations at second order. 
It will be also shown that there is a striking connection between 
this algebraic structure 
and the sign of the energy at lowest order in the deformation. 
These results will follow from a more general classification theorem
that will be proven here in $n\ge 3$ dimensions for 
nonlinear \grav/ gauge theories 
with more derivatives than contained in the linear theory
but which preserve the number of local dynamical degrees of freedom. 
For this analysis, 
it turns out to be most convenient to 
employ a linearized spin-connection/frame
formalism for free \grav/ fields,
in terms of which the deformation obtained in $n=3$ dimensions 
in \Ref{exoticth}
takes its simplest form 
(indeed, no other complete formulation for that deformation 
has yet been derived).

\section{ Deformation analysis and Main results }

We introduce the set of $N\ge 1$ fields 
$\h{a\mu}{}{A}$ (viewed as linearized frames), 
together with a set of auxiliary fields 
(with the role of linearized spin-connections)
\EQ
\wtop{1}{a\mu\nu}{}{A} = 
\der{a}\h{[bc]}{}{A} \sodder{b}{\mu}\sodder{c}{\nu}
-2\sodder{b}{[\nu} \der{\mu]}\h{(ab)}{}{A} , 
\label{weq}
\doneEQ
in terms of $\h{ab}{}{A}= \invsodder{b}{\mu}\h{a\mu}{}{A}$, 
${\internal A}=1,\ldots,N$, 
where $\invsodder{b}{\mu}$ denotes any fixed orthonormal frame 
for the Minkowski metric 
$\flat{ab}=\invsodder{a}{\mu}\invsodder{b}{\nu} \flat{\mu\nu}$
(with $\invflat{\mu\nu}= {\rm diag}(-1,+1,\ldots,+1)$)
while $\sodder{b}{\mu}$ is the inverse frame
\cite{indices}. 
The free \grav/ theory for these fields 
on $n$-dimensional Minkowski space $(\Rnum^n,\flat{ab})$
is given by the linear field equations
\EQ
\Etop{1}{a\mu}{}{A} = 
\Rtop{1}{ab\mu\nu}{}{A} \sodder{b\nu}{} =0 ,\quad
\Rtop{1}{ab\mu\nu}{}{A} = \der{[a} \wtop{1}{b]\mu\nu}{}{A} , 
\label{linheq}
\doneEQ
and the abelian gauge symmetries
\EQs
\varh{\xi}{0}{a\mu}{}{A} = \der{a} \vect{\mu}{}{A} ,\quad
\varh{\chi}{0}{a\mu}{}{A} = \invsodder{a}{\nu} \spin{\nu\mu}{}{A} 
\label{abelianhvar}
\doneEQs
involving parameters 
$\vect{\mu}{}{A},\spin{\nu\mu}{}{A}=\spin{[\nu\mu]}{}{A}$
which are arbitrary functions of the spacetime coordinates. 
Here $\Rtop{1}{ab\mu\nu}{}{A}$ is a gauge invariant field strength,
as is seen due to 
\EQ
\varwtop{\xi}{0}{1}{a\mu\nu}{}{A} = 0 ,\quad
\varwtop{\chi}{0}{1}{a\mu\nu}{}{A} = \der{a} \spin{\nu\mu}{}{A} .
\doneEQ
This free theory comes from the gauge-invariant Lagrangian
\EQ
\Ltop{2} =
-\tfrac{1}{2}( 
3\h{[a}{\mu}{A} \der{b}\wtop{1}{c]}{\nu\rho}{B} \sodder{c}{\rho}
+ \wtop{1}{[a}{\mu\rho}{A} \wtop{1}{b]\rho}{\nu}{B} )
\sodder{a}{\mu}\sodder{b}{\nu} \id{AB}{} 
\label{quadrL}
\doneEQ
where $\id{AB}{}$ is a fixed symmetric matrix. 
A variation of $\h{a\mu}{}{A}$ yields the field equations \eqref{linheq}
to within irrelevant trace terms,
\EQ
\id{}{AB} 
( \id{a}{b}\id{\mu}{\nu} +\tfrac{1}{2-n} \sodder{b\nu}{} \invsodder{a\mu}{} )
\delta\Ltop{2}/\delta\h{b\nu}{}{B} 
= \Etop{1}{a\mu}{}{A} , 
\doneEQ
while an independent variation of $\wtop{1}{a\mu\nu}{}{A}$ gives 
a total divergence as a consequence of the auxiliary equations \eqref{weq}.
Hence it is necessary only to vary $\h{a\mu}{}{A}$ alone
in considering variations of the Lagrangian \eqref{quadrL}
(analogous to a ``1.5 formalism'' \cite{supergravity}). 
To see how the formalism here is related to 
the familiar free theory of \grav/ fields, observe that 
if the gauge condition $\h{[ab]}{}{A}=0$ is imposed
using the gauge freedom 
$\h{ab}{}{A} \rightarrow 
\h{ab}{}{A} + \spin{ab}{}{A}$
involving the skew-tensor functions 
$\spin{ab}{}{A}=\invsodder{a}{\nu}\invsodder{b}{\rho}\spin{\nu\rho}{}{A}$,
then the field equations \eqref{linheq} 
and gauge symmetries \eqref{abelianhvar}
reduce to the ordinary Fierz-Pauli \grav/ equations 
\EQ
\invflat{cd} \der{[a|}\der{[b} \symmh{c]|d]}{}{A} =0
\label{freefieldeq}
\doneEQ
and \grav/ gauge invariance 
\EQ
\symmh{cd}{}{A} \rightarrow \symmh{cd}{}{A} +\der{(c}\vect{d)}{}{A}
\doneEQ
in terms of the symmetric tensor fields 
$\symmh{cd}{}{A} = \h{(cd)}{}{A}$
and covector functions $\vect{b}{}{A} = \invsodder{b}{\mu} \vect{\mu}{}{A}$.
Furthermore, note these \grav/ fields
will have positive energy as obtained from 
the conserved stress-energy tensor of the free Lagrangian \eqref{quadrL}
if (and only if) $\id{AB}{}=diag(+1,\ldots,+1)$ 
is a positive definite matrix.

We now consider deformations of the gauge symmetries and field equations
\EQs
&&
\delta_{\xi}\h{a\mu}{}{A} =
\varh{\xi}{0}{a\mu}{}{A} + \varh{\xi}{1}{a\mu}{}{A} + \cdots ,\quad
\delta_{\chi}\h{a\mu}{}{A} =
\varh{\chi}{0}{a\mu}{}{A} + \varh{\chi}{1}{a\mu}{}{A} + \cdots , 
\label{deformhvar}\\
&&
\E{a\mu}{}{A} 
= \Etop{1}{a\mu}{}{A} + \Etop{2}{a\mu}{}{A} + \cdots 
= \ELop{a\mu}{}{A}{L}, \quad
L= \Ltop{2} + \Ltop{3} + \cdots , 
\label{deformheq}
\doneEQs
satisfying gauge invariance of the Lagrangian $L$,
so that $\delta_\xi L$ and $\delta_\chi L$ are total divergences. 
This determining condition has the following direct formulation
(see \Ref{AMSpaper,Annalspaper})
\EQ
\ELop{a\mu}{}{A}{ \delta_{\xi}\h{b\nu}{}{B}\coE{b\nu}{}{B} }
= \ELop{a\mu}{}{A}{ \delta_{\chi}\h{b\nu}{}{B}\coE{b\nu}{}{B} }
=0
\label{gaugeinv}
\doneEQ
where $\ELop{a\mu}{}{A}{\cdot}$ is the Euler-Lagrange operator 
with respect to $\h{a\mu}{}{A}$. 
Two deformations will be regarded as equivalent 
if they are related by field redefinitions
\EQ
\h{a\mu}{}{A} \rightarrow \newh{}{a\mu}{}{A} 
= \h{a\mu}{}{A} +\newh{(2)}{a\mu}{}{A} +\cdots
\doneEQ
or by parameter redefinitions
\EQ
\vect{\mu}{}{A} \rightarrow \newvect{}{\mu}{}{A} 
= \vect{\mu}{}{A} +\newvect{(1)}{\mu}{}{A} +\cdots ,\quad
\spin{\mu\nu}{}{A} \rightarrow \newspin{}{\mu\nu}{}{A} 
= \spin{\mu\nu}{}{A} +\newspin{(1)}{\mu\nu}{}{A} +\cdots .
\doneEQ
The deformation terms are taken to be locally constructed 
from the Minkowski frame, metric and volume tensors, 
\grav/ fields, gauge symmetry parameters, 
and derivatives of the fields and parameters. 
As the main assumption,
the derivatives appearing in the deformation terms 
will be restricted so that 
the number of local dynamical degrees of freedom of the \grav/ fields 
in the linear theory 
is preserved order by order in a nonlinear deformation.
With this requirement
the most general possible form for nonlinear \grav/ field equations
is that of a quasilinear second order system of PDEs, 
namely, highest derivative terms are of second order,
while the coefficient of these terms depends on at most 
first order derivatives of the \grav/ fields.
In turn, due to a general relation known to hold 
\cite{AMSpaper,Deser}
between the form of lowest order deformation terms in the field equations
and Noether currents of rigid symmetries associated with 
the lowest order deformation terms in the gauge symmetries,
the most general possible form for \grav/ gauge symmetries
is required to be at most first order in derivatives. 
It is worth noting that 
stronger assumptions have been made in all systematic analyses to-date 
and essentially lead to nonlinear \grav/ field equations
being restricted to the form of a second order system of semilinear PDEs,
where the coefficient of the second order derivative terms involves 
{\it no} derivatives of the \grav/ field
but is allowed to depend on the \grav/ field itself.
Such a form arises directly if the deformation terms in the Lagrangian
are restricted to contain at most two derivatives, \ie/
second order derivatives appear linearly, 
or more generally, first order derivatives appear quadratically.
In contrast, 
the weaker assumptions made here 
allow these deformation terms to have a general polynomial depedence
on first derivatives 
(with no higher order derivatives appearing),
which is compatible with a quasilinear form for the \grav/ field equations
as occurs for the deformations investigated in \Ref{exoticth}. 

In addition, 
corresponding to the role of the auxiliary field \eqref{weq} 
in the free theory,
all derivatives of the \grav/ fields $\h{a\mu}{}{A}$
in a nonlinear deformation will be assumed to appear only through 
$\wtop{1}{a\mu\nu}{}{A}$. 
Consequently it follows that at lowest order the deformation of 
the gauge symmetries and field equations 
is required to take the form (indices suppressed)
\EQs
&&
\var{1}{\chi} h = A h \chi + B w \chi + C h \der{} \chi ,\quad
\var{1}{\xi} h = F h \xi + G w \xi + H h \der{} \xi , 
\label{linvarh}
\doneEQs
and 
\EQ
\Etop{2}{}{}{} = I w\der{} w + J h\der{} w + K w w +M h w +N h h , 
\label{quadrheq}
\doneEQ
where the coefficients are constant tensors $A,\ldots,N$ 
locally constructed just from 
$\invsodder{a}{\mu}$, $\flat{ab}$, $\vol{a_1\cdots a_n}$. 
Such deformations will be called ``quasilinear covariant''. 

A useful field-theoretic formulation of gauge invariance \eqref{gaugeinv}
is given by the following necessary and sufficient 
Lie derivative equations
(indices suppressed)
\EQs
&& 
\Lie{\delta_\xi} E =0 ,\quad 
\Lie{\delta_\chi} E =0 ,
\label{liedereq}\\
&&
\Lie{[\delta_{\xi_1},\delta_{\xi_2}]} E =0 ,\quad
\Lie{[\delta_{\chi_1},\delta_{\chi_2}]} E=0 ,\quad
\Lie{[\delta_{\xi_1},\delta_{\chi_1}]} E =0 , 
\label{liedercommeq}
\doneEQs
where $\Lie{\delta}$ denotes the Lie derivative operator 
as defined with respect to field variations $\delta\h{}{}{}$
regarded formally as tangent vector fields 
on the space of field configurations $\h{}{}{}$
(see \Ref{multigravpaper}).
An expansion of these equations \eqrefs{liedereq}{liedercommeq}
in powers of $\h{}{}{}$ 
gives a system of determining equations 
for all allowed deformation terms 
order by order in the field equations and gauge symmetries. 

The first order deformation terms \eqrefs{linvarh}{quadrheq}
can be determined by solving the expanded determining equations
\eqref{liedercommeq} to zeroth order
and \eqref{liedereq} to first order
using the methods of \Ref{Annalspaper}.
This leads to the following classification result.

\Proclaim{Theorem 1.}{
All first order quasilinear covariant deformations \eqrefs{linvarh}{quadrheq}
of the free \grav/ gauge theory \eqsref{weq}{quadrL}
in $n\ge 3$ dimensions 
are equivalent to a combination of the types 
\begin{subequations}
\label{gravitytype}
\EQs
&&
\varh{\xi}{1}{a\mu}{}{A} 
= \a{A}{BC} \wtop{1}{a\mu\nu}{}{B} \vect{}{\nu}{C} ,\quad
\varh{\chi}{1}{a\mu}{}{A} 
= \a{A}{BC} \h{a}{\nu}{B} \spin{\nu\mu}{}{C} ,
\\
&& 
\Ltop{3} =
-(\a{}{ABC}( 
3\Rtop{1}{[ab}{\alpha\beta}{A} \h{c}{\nu}{B} \h{d]}{\rho}{C} \sodder{d}{\rho}
+\tfrac{3}{2} \Rtop{2}{[ab}{\alpha\beta}{A} \h{c]}{\nu}{B} )
\sodder{c}{\nu}
+\id{AB}{} \tfrac{1}{2} \wtop{2}{[a}{\alpha\rho}{A} \wtop{1}{b]\rho}{\beta}{B} 
)\sodder{a}{\alpha} \sodder{b}{\beta} ,
\nonumber\\
\\
&&
\a{}{ABC}=\a{}{(ABC)}
\label{aeq}
\doneEQs
\end{subequations}
where
\EQs
&&
\Rtop{2}{ab}{\nu\rho}{A} =
\der{[a} \wtop{2}{b]}{\nu\rho}{A} 
+\a{A}{BC} \wtop{1}{[a}{\nu\mu}{B}\wtop{1}{b]\mu}{\rho}{C} ,
\\
&&
\wtop{2}{a}{\nu\rho}{A} =
\a{A}{BC}( 
( 2\h{}{b[\nu}{B} \sodder{\rho]c}{} \invsodder{a\mu}{}
-\h{a\mu}{}{B} \sodder{b\nu}{}\sodder{c\rho}{} ) \der{[b}\h{c]}{\mu}{C}
-2\h{}{b[\nu}{B} \der{[a}\h{b]}{\rho]}{C} )
\doneEQs
(corresponding to an algebra-valued gravitational interaction),
or if $n=3$
\begin{subequations}
\label{3dtype}
\EQs
&&
\varh{\xi}{1}{a\mu}{}{A} = 0 ,\quad
\varh{\chi}{1}{a\mu}{}{A} 
= \b{A}{BC} \invvol{\nu\rho\alpha} \wtop{1}{a\nu\rho}{}{B} 
\spin{\alpha\mu}{}{C} ,
\\
&& \Ltop{3} =
\tfrac{1}{2} \b{}{ABC} \invvol{abc}
\wtop{1}{a}{\alpha\beta}{A} \wtop{1}{b\alpha}{\nu}{B}\wtop{1}{c\nu\beta}{}{C} ,
\\
&&
\b{}{ABC}=\b{}{(ABC)}
\label{beq}
\doneEQs
\end{subequations}
(corresponding to the parity violating part of 
an algebra-valued topological gravity interaction),
or if $n=5$
\begin{subequations}
\label{5dtype}
\EQs
&&
\varh{\xi}{1}{a\mu}{}{A} = 0 ,\quad
\varh{\chi}{1}{a\mu}{}{A} 
= \c{A}{BC} \vol{\mu\nu\rho\alpha\beta} \wtop{1}{a}{\nu\rho}{B} 
\spin{}{\alpha\beta}{C} ,
\label{linvarhspin5d}
\\
&& \Ltop{3} =
-\tfrac{1}{2} \c{}{ABC} \invvol{abc\nu\rho}
\wtop{1}{a}{\alpha\beta}{A}(
\wtop{1}{b\alpha\beta}{}{B}\wtop{1}{c\nu\rho}{}{C}
- 4\wtop{1}{b\alpha\nu}{}{B}\wtop{1}{c\beta\rho}{}{C} ) ,
\label{cubicL5d}
\\
&&
\c{}{ABC}=\c{}{[ABC]}
\label{ceq}
\doneEQs
\end{subequations}
(corresponding to a parity violating exotic interaction). 
The structure on the internal vector space $\Rnum^N$ 
associated with the set of $N\geq 1$ \grav/ fields $\h{a\mu}{}{A}$ 
is a commutative, invariant-normed algebra 
in the first two types of deformations
and an anticommutative, invariant-normed algebra 
in the third type of deformation. 
}

These results determine the commutator structure of
the deformed gauge symmetries to lowest order, 
$\varcomm{0}{1}{2} =\var{0}{3}$.
For the type \eqref{gravitytype},
the nonvanishing commutators are given by 
\EQ
\varcomm{0}{\chi_1}{\chi_2} =\var{0}{\chi_3}
\doneEQ
with 
\EQ
\spinsub{3}{\mu\nu}{}{A} 
= 2\spinsub{1}{[\mu}{\rho}{B} \spinsub{2}{\nu]\rho}{}{C} \a{A}{BC} ,
\doneEQ
and 
\EQ
\varcomm{0}{\chi_1}{\xi_1} =\var{0}{\xi_3}
\doneEQ
with 
\EQ
\vectsub{3}{\mu}{}{A} 
= \spinsub{1}{\mu\nu}{}{B} \vectsub{1}{}{\nu}{C} \a{A}{BC} .
\doneEQ
In contrast,
the only nonvanishing commutator for the types \eqrefs{3dtype}{5dtype} 
is given by 
\EQ
\varcomm{0}{\chi_1}{\chi_2} =\var{0}{\xi_3}
\doneEQ
with, respectively,
\EQ
\vectsub{3}{\mu}{}{A} 
= \invvol{\nu\rho\alpha} \spinsub{1}{\nu\rho}{}{B} 
\spinsub{2}{\alpha\mu}{}{C} \b{A}{BC} 
\text{ when $n=3$ }
\doneEQ
and
\EQ
\vectsub{3}{\mu}{}{A} 
= \vol{\mu\nu\rho\alpha\beta} \spinsub{1}{}{\nu\rho}{B} 
\spinsub{2}{}{\alpha\beta}{C} \c{A}{BC}
\text{ when $n=5$ }. 
\doneEQ
An analysis of the determining equation \eqref{liedercommeq}
by the methods of \Ref{Annalspaper}
shows that the same commutator structure holds at next order
\EQ
\varcomm{1}{1}{2}\onshell =\var{1}{3}\onshell
\label{lincommeq}
\doneEQ
when $\h{a\mu}{}{A}$ satisfies the linear field equation \eqref{linheq}. 

Higher order deformation terms
in the gauge symmetries and field equations
can be derived by continuing to solve 
the determining equations \eqrefs{liedereq}{liedercommeq}
at successively higher orders. 
However, an integrability condition 
on the first order deformation terms \eqsref{gravitytype}{5dtype}
arises from the closure result for the gauge symmetry commutator structure
at first order in \eqref{lincommeq}
if we consider the terms that involve second derivatives of $\h{a\mu}{}{A}$.
For the type \eqref{gravitytype}, 
all such terms come from the commutator 
$[\var{1}{\xi_1},\var{1}{\xi_2}] \symmh{ab}{}{A}$,
which yields
\EQ
\a{A}{B[C}\a{B}{D]E} \vectsub{1}{}{e}{D} \vectsub{2}{}{d}{C} 
\Rtop{1}{a(ed)b}{}{E}\onshell .
\doneEQ
These terms must vanish to within a symmetrized derivative 
by \eqref{lincommeq}. 
Since $\der{[g|}\der{[f}\Rtop{1}{a](de)|b]}{}{E} \neq 0$
we consequently obtain the integrability condition 
\EQ
\a{A}{B[C}\a{B}{D]E} =0 . 
\label{aaeq}
\doneEQ
Next, for the type \eqref{3dtype}
we find the terms that come from 
$[\var{1}{\chi_1},\var{1}{\chi_2}] \symmh{ab}{}{A}$
with second derivatives of $\h{a\mu}{}{A}$
are given by 
\EQ
\b{A}{B[C}\b{B}{D]E} \tspinsub{1}{}{e}{D} \tspinsub{2}{}{d}{C} 
\Rtop{1}{a(ed)b}{}{E}\onshell
\label{3dobstruct}
\doneEQ
where $\tspin{}{\nu}{A} = \invvol{\nu\alpha\beta} \spin{\alpha\beta}{}{A}$.
But \eqref{3dobstruct} vanishes since, in $n=3$ dimensions, 
the linear \grav/ field equation \eqref{linheq} is well known to imply
$\Rtop{1}{adeb}{}{E}=0$. 
Thus, due to the absence of local dynamical degrees of freedom, 
no integrability condition arises from equation \eqref{lincommeq} 
for the deformation \eqref{3dtype}.
(However, if the deformations \eqrefs{gravitytype}{3dtype} 
are combined in $n=3$ dimensions, 
then we obtain an integrability condition
\EQ
\a{A}{BC}\b{B}{DE} = \b{A}{BE}\a{B}{DC} .)
\label{abeq}
\doneEQ 
Finally, for the type \eqref{5dtype}
the same commutator now yields 
\EQ
\c{A}{BC}\c{B}{DE} ( \spinsub{1}{}{de}{C} \spinsub{2}{(a}{c}{D} 
- \spinsub{1}{(a}{c}{D} \spinsub{2}{}{de}{C} )
\Rtop{1}{b)cde}{}{E} \onshell
+\flat{ab}\ \eqtext{terms}
\label{5dobstruct}
\doneEQ
which does not vanish to within a symmetrized derivative. 
Hence, it follows from equation \eqref{lincommeq} that 
these terms \eqref{5dobstruct} must be canceled by 
suitable quadratic deformation terms of the form (indices suppressed)
\EQ
\var{2}{\chi} h 
= bb \ontop{(1)}{R} h \chi + \eqtext{lower\ derivative\ terms} . 
\label{quadrvarhspin}
\doneEQ
In turn, a similar analysis of the resulting commutator
$[\var{1}{\xi_1},\var{1}{\chi_1}] \symmh{ab}{}{A}$
leads to further quadratic deformation terms 
\EQ
\var{2}{\xi} h 
= bb \ontop{(1)}{R} h \der{}\xi + \eqtext{lower\ derivative\ terms} . 
\label{quadrvarhvect}
\doneEQ
Then we find that the commutator
$[\var{1}{\xi_1},\var{1}{\xi_2}] \symmh{ab}{}{A}$
produces second derivative terms of the same form as \eqref{5dobstruct}
where 
$\spinsub{1}{de}{}{D}$ and $\spinsub{2}{ac}{}{C}$
are replaced by 
$\der{d}\vectsub{1}{e}{}{D}$ and $\der{a}\vectsub{1}{c}{}{C}$.
Since the resulting terms do not vanish to within a symmetrized derivative,
we thus derive an integrability condition
\EQ
\c{A}{BC}\c{B}{DE} =0 . 
\label{cceq}
\doneEQ
(Similar integrability conditions occur if the deformations
\eqrefs{5dtype}{gravitytype} are combined in $n=5$ dimensions, 
\EQ
\a{A}{BC} \c{B}{DE} =\c{A}{BC} \a{B}{DE} =  0 .)
\label{aceq}
\doneEQ 

The integrability conditions \eqrefs{aaeq}{cceq}
assert that the underlying internal algebras associated with the \grav/ fields
in the deformations \eqrefs{gravitytype}{5dtype}
are, respectively, associative and nilpotent of degree three. 
By the results in \Ref{Henneaux1,exoticth},
there are no further integrability conditions 
on the construction of deformations \eqrefs{gravitytype}{3dtype}
to all higher orders in solving the determining equations. 
On the other hand, 
the deformation \eqref{5dtype} 
can be shown to satisfy the determining equations to all orders itself, 
since any higher order deformation terms necessarily vanish 
as a consequence of the nilpotency \eqref{cceq} of the algebra. 
Thus, this deformation 
\EQ
L=\Ltop{2}+\Ltop{3} ,\quad
\delta_{\xi}\h{a\mu}{}{A} = \varh{\xi}{0}{a\mu}{}{A} ,\quad
\delta_{\chi}\h{a\mu}{}{A} 
= \varh{\chi}{0}{a\mu}{}{A} + \varh{\chi}{1}{a\mu}{}{A} 
\label{5dL}
\doneEQ
given by \eqref{abelianhvar}, \eqref{quadrL}, \eqrefs{5dtype}{aaeq}
yields a full, nonlinear \grav/ gauge theory. 
Hence we arrive at the following main classification result.

\Proclaim{ Theorem 2.}{
The nonlinear \grav/ gauge theories in $n>2$ dimensions 
determined by the respective first-order deformations 
\eqref{gravitytype}, \eqref{3dtype}, \eqref{5dtype} 
are equivalent to an algebra-valued Einstein gravity theory for $n\ge 3$
with a commutative, associative, invariant-normed algebra, 
or if $n=3$, 
a novel nonlinear theory related to a scaling limit of
algebra-valued topological gravity theory 
with a commutative, invariant-normed algebra, 
or if $n=5$
a new algebra-valued nonlinear theory 
with an anticommutative, nilpotent, invariant-normed algebra. 
Additional nonlinear \grav/ gauge theories arise from 
the gravity deformation \eqref{gravitytype} combined with either of 
the other two deformations \eqref{3dtype} or \eqref{5dtype},
describing exotic (parity violating) generalizations of 
algebra-valued Einstein gravity theory
in $n=3,5$ dimensions
(with the algebras restricted by conditions \eqrefs{abeq}{aceq}). 
There are no other nonlinear \grav/ gauge theories of 
quasilinear covariant type. 
}

The five-dimensional nonlinear theory without gravitational interactions 
has the following features. 
Its field equations (to within trace terms) 
\EQ
\E{a\mu}{}{A} = \R{a\mu}{}{A}=\der{[a}\w{b]\mu\nu}{}{A} \sodder{b\nu}{}=0
\doneEQ
are given by the quadratic spin-connection
\EQ
\w{a\mu\nu}{}{A} = 
\wtop{1}{a\mu\nu}{}{A} + \W{a\mu\nu}{}{A} 
-2\invsodder{a}{\rho}\sodder{b}{[\mu|}\W{b|\nu]\rho}{}{A}
+\tfrac{4}{3}\sodder{b\rho}{}\invsodder{a[\mu|}{} \W{b\rho|\nu]}{}{A}
\doneEQ
with 
\EQ
\W{}{a\alpha\beta}{A} = 
\c{A}{BC}( 
2\invvol{abc\nu\rho}( 2\wtop{1}{b\nu}{\alpha}{B}\wtop{1}{c\rho}{\beta}{C} 
-\wtop{1}{b}{\alpha\beta}{B}\wtop{1}{c\nu\rho}{}{C} )
-\invvol{abc\alpha\beta} \wtop{1}{b}{\nu\rho}{B}\wtop{1}{c\nu\rho}{}{C} 
+2\invvol{abc\rho[\alpha} \wtop{1}{b}{\beta]\mu}{B}\wtop{1}{c\mu\rho}{}{C} )
\doneEQ
where 
$\W{a}{\alpha\beta}{A}
- \tfrac{2}{3}\W{c}{\rho[\beta}{A} \invsodder{a}{\alpha]}\sodder{c}{\rho}$
plays the role of a contorsion tensor. 
Exponentiation of its gauge symmetries 
generates a group of finite gauge transformations 
$\h{a\mu}{}{A} \rightarrow \newh{}{a\mu}{}{A}$ given by 
\EQ
\newh{}{a\mu}{}{A} = 
\h{a\mu}{}{A} + \der{a} \vect{\mu}{}{A} 
+ \invsodder{a}{\nu} \spin{\nu\mu}{}{A} 
+ \c{A}{BC} \vol{\mu\nu\rho\alpha\beta} (
\wtop{1}{a}{\nu\rho}{B} +\tfrac{1}{2} \der{a} \spin{}{\nu\rho}{A} )
\spin{}{\alpha\beta}{C} . 
\doneEQ
Using this gauge freedom in the theory, 
we can make a nonlinear change of field variable to 
$\newsymmh{ab}{}{A} = \invsodder{(a}{\mu}\newh{}{b)\mu}{}{A}$
with $\spin{\nu\mu}{}{A}$ determined in terms of $\h{a\mu}{}{A}$
by $\invsodder{[a}{\mu}\newh{}{b]\mu}{}{A}=0$. 
The gauge symmetries then consist of 
(after a parameter redefinition)
\EQ
\label{5dgaugesymm}
\delta_\xi \newsymmh{ab}{}{A} = 
\der{(a} \vect{b)}{}{A} 
+ \c{A}{BC} \cross{(a|}{cdef} \der{c} \newsymmh{d|b)}{}{B} 
\der{e}\vect{f}{}{C} , 
\doneEQ
while the field equations become
\EQ
\label{5dfieldeq}
\newE{ab}{}{A} = 
-\invflat{cd}(
2\der{[a|}\der{[b}\newsymmh{c]|d]}{}{A} 
+ \der{c}( \newW{(ab)d}{}{A}(\newsymmh{}{}{})
+\tfrac{1}{3} \newW{ed}{e}{A}(\newsymmh{}{}{})\flat{ab} ) 
)
=0
\doneEQ
which are of quasilinear second order form,
where
\EQ
\newW{amn}{}{A}(\newsymmh{}{}{}) = 
4\c{A}{BC}( 
\vol{abcde} \curlh{m}{db}{}{B}\curlh{n}{ec}{}{C}
- \vol{abcmn} \curlh{}{bde}{}{B}\curlh{}{c}{de}{C}
- \vol{abcd[m} \curlh{}{b}{n]e}{B}\curlh{}{edc}{}{C})
\doneEQ
with $\curlh{amn}{}{}{A} = \der{[m}\newsymmh{n]a}{}{A}$. 

Since the algebra $(\c{A}{BC},\Rnum^N)$
associated with the \grav/ fields in this theory 
is anticommutative and nilpotent of degree three, 
it satisfies the Jacobi identity and hence is equivalently characterized
as being a solvable Lie algebra of length two \cite{liealg},
with an invariant norm. 
Note that the existence of an invariant norm puts some restriction 
on the Lie bracket structure in the algebra.  
The simplest example of such an algebra  $(\c{A}{BC},\Rnum^N)$
is given by 
$\c{}{ABC}=u\downindex{[A}v\downindex{B}w\downindex{C]}$
where $u\downindex{A},v\downindex{B},w\downindex{C}$
are mutually orthogonal null vectors in a $N=6$ dimensional vector space
with norm $\id{AB}{}=diag(+1,+1,+1,-1,-1,-1)$.

However, if we impose a physically natural requirement that
the individual \grav/ fields should have positive energy
(or more precisely that the weak energy condition \cite{Wald-book} must hold),
using the conserved stress-energy tensor derived from the Lagrangian
\eqref{5dL}, 
this severely restricts the allowed nonabelian structure of the algebra. 
Specifically, 
as already noted for the free theory, 
positivity of energy forces the norm on the algebra to be positive definite. 
As a consequence, 
from nilpotency \eqref{cceq} combined with norm-invariance \eqref{ceq},
we have $\c{A}{BC}\c{}{ADE} =0$,
which implies $\c{A}{BC}=0$ due to positive definiteness of the norm. 
Hence, in this situation, 
every anticommutative, nilpotent, invariant-normed algebra 
$(\c{A}{BC},\Rnum^N)$ is abelian. 
In a similar manner, 
as shown in \Ref{Henneaux1},
any commutative, associative, invariant-normed algebra 
$(\a{A}{BC},\Rnum^N)$ 
with a positive definite norm 
is the direct sum of one-dimensional unit algebras $\Rnum$. 
Thus we obtain the following no-go result. 

\Proclaim{ Theorem 3.}{
The only quasilinear covariant deformations
in $n\ge 4$ dimensions for \grav/ gauge theories with positive energy 
are semilinear, 
in particular, equivalent to Einstein gravity theory
with no interaction between different \grav/ fields. }

This strengthens the no-go theorem in \Ref{Henneaux1} 
to the more general class of quasilinear covariant deformations 
considered here. 
In $n=3$ dimensions, 
positivity of energy is compatible with 
a nontrivial algebra structure $(\b{A}{BC},\Rnum^N)$,
as emphasized in \Ref{exoticth}. 
The resulting three-dimensional nonlinear theory, using the formalism here,
without gravitational interactions
is given by the gauge symmetries
\EQ
\delta_{\xi} \h{a\mu}{}{A} = \der{a} \vect{\mu}{}{A}, \quad
\delta_{\chi} \h{a\mu}{}{A} = 
\vol{a\mu\nu} \tspin{}{\nu}{A} 
+ \b{A}{BC} \vol{\mu\nu\rho} \tw{a}{\nu}{B} \tspin{}{\rho}{C} ,
\doneEQ
and the field equations
\EQ
\E{a\mu}{}{A} = \der{[a} \tw{b]}{\nu}{A} \cross{\mu\nu}{b}
\doneEQ
where $\tw{a}{\nu}{A}=\invvol{\nu\mu\rho} \w{a\mu\rho}{}{A}$
satisfies the following quadratic spin-connection equation
\EQ
\der{[a}\h{b]\mu}{}{A}
= \vol{\mu\nu\rho}( \invsodder{[a}{\nu} \tw{b]}{\rho}{A}
+ \b{A}{BC} \tw{[a}{\nu}{B} \tw{b]}{\rho}{C} ) . 
\label{3dweq}
\doneEQ
It is possible to solve \eqref{3dweq} to obtain 
\EQ
\tw{a\nu}{}{A}
=\powW{a\nu}{}{A}{1/2}-\tfrac{1}{2} \invsodder{a\nu}{} \unit{A}
\doneEQ
in terms of the square root of 
\EQ
\W{a\nu}{}{A}= 
-2\cross{a}{bc} \der{b}\h{c\nu}{}{A} +\tfrac{1}{2} \invsodder{a\nu}{} \unit{A}
\doneEQ
as defined by 
\EQ
\cross{c}{ab} \cross{\nu}{\mu\rho}
\powW{a\mu}{}{B}{1/2}\powW{b\rho}{}{C}{1/2}
= - \W{c\nu}{}{A}
\doneEQ
where $\unit{A}$ is a unit element (appended if necessary) 
in the algebra $(\b{A}{BC},\Rnum^N)$. 
(Note, with indices suppressed, 
this square root satisfies the algebraic relation
$\powW{}{}{}{1/2} \times \powW{}{}{}{1/2} =-\W{}{}{}$
with $\times$ being a symmetric product 
on $\Rnum^3\otimes\Rnum^3\otimes\Rnum^N$ given by 
the tensor product of three-dimensional cross-products $\vol{}$
combined with the algebra product $\b{}{}$.)
This theory can be formulated in terms of ordinary \grav/ fields
$\newsymmh{ab}{}{A} = \newh{}{(ab)}{}{A}$
analogously to the five-dimensional theory
by a suitable field redefinition using the finite gauge transformations
\EQ
\newh{}{a\mu}{}{A} = 
\h{a\mu}{}{A} + \der{a} \vect{\mu}{}{A} + \vol{a\mu\nu} \tspin{}{\nu}{A} 
+ \b{A}{BC} \vol{\mu\nu\rho}(
\tfrac{1}{2}\der{a}\tspin{}{\nu}{B} +  \tw{a}{\nu}{B} ) \tspin{}{\rho}{C}
\doneEQ
generated from the gauge symmetries on solutions of the field equations.

\section{ Concluding remarks }

It is worth investigating to what extent
the complete five-dimensional nonlinear \grav/ theory derived here 
is sensible as a classical field theory 
in view of its unusual features. 

First, the field equations \eqref{5dfieldeq} possess 
a well-posed initial value formulation. 
The linear part of \eqref{5dfieldeq} is given by 
the free Fierz-Pauli \grav/ equations \eqref{freefieldeq},
which are a second order hyperbolic system
whose characteristic directions coincide with the Minkowski light cones, 
when suitable gauge conditions are imposed. 
While the nonlinear part of \eqref{5dfieldeq} 
also involves second order derivatives 
and a priori might be expected to alter the characteristic directions,
in fact the nilpotency of the internal algebra $(\c{A}{BC},\Rnum^N) =\alg$ 
implies that the full field equations are a decoupled triangular system of
semilinear Fierz-Pauli equations. 
In particular, consider $\Dalg= [\alg,\alg]$
where the bracket denotes the algebra product $\c{A}{BC}$ 
on the internal normed vector space $(\Rnum^N, \id{AB}{})$. 
Due to the nilpotency and indefinite sign of the norm of $\alg$, 
$\Dalg$ is a null, abelian subalgebra of $\alg$, 
with a null complement $\coDalg$ 
such that $\Dalg\oplus \coDalg=\alg$
(once we divide out by any trivial abelian factors of $\alg$). 
Since $\Dalg$ and $\coDalg$ obey 
$[\coDalg,\coDalg] \subseteq \Dalg$
and $[\Dalg,\Dalg]=[\Dalg,\coDalg]=0$, 
the subset of \grav/ fields associated with $\coDalg$
(\ie/ $\PcoDalg(\h{a\mu}{}{A})$ where 
$\PcoDalg=\PTDalg$ is the transpose of the projector $\PDalg$ onto $\Dalg$)
satisfies free Fierz-Pauli equations \eqref{freefieldeq},
and the remaining subset of \grav/ fields associated with $\Dalg$
(\ie/ $\PDalg(\h{a\mu}{}{A})$)
satisfies inhomogeneous Fierz-Pauli equations with 
quadratic source terms that involve only the free \grav/ fields. 
For example, 
consider the case of the algebra given by 
$\c{}{ABC}=u\downindex{[A}v\downindex{B}w\downindex{C]}$
on $(\Rnum^6,\id{AB}{}=
2 u\downindex{(A}u'\downindex{B)}+ 2 v\downindex{(A}v'\downindex{B)}
+2 w\downindex{(A}w'\downindex{B)})$
with a null vector basis 
$u\downindex{A},v\downindex{B},w\downindex{C}, 
u'\downindex{A},v'\downindex{B},w'\downindex{C}$
(whose only non-zero inner products are 
$u'\downindex{A} u\upindex{A}
= v'\downindex{A} v\upindex{A}
=w'\downindex{A} w\upindex{A}= 1$). 
The inhomogeneous Fierz-Pauli equations hold for the fields
$\P{A}{B}\h{a\mu}{}{B}$
given by the null projector 
$\P{A}{B}=
u\upindex{A}u'\downindex{B}+ v\upindex{A}v'\downindex{B}
+ w\upindex{A}w'\downindex{B}$,
while the free \grav/ Fierz-Pauli equations
involve the fields 
$\PT{A}{B}\h{a\mu}{}{B}$
given in terms of the transpose null projector 
$\PT{A}{B}=\id{BC}{} \id{}{AD} \P{C}{D}
= u'\upindex{A}u\downindex{B}+ v'\upindex{A}v\downindex{B}
+ w'\upindex{A}w\downindex{B}$. 

As a consequence of this decoupling feature, 
the well-posedness property of the field equations \eqref{5dfieldeq}
is insensitive to the lack of positivity of energy \cite{energy}
arising from the conserved stress-energy tensor of the Lagrangian \eqref{5dL}
through the indefinite sign of the norm on the algebra $\alg$. 

Finally, although the theory exists only in five dimensions,
it is relevant for four dimensions if a Kaluza-Klein reduction is considered. 
We begin from a product decomposition of 
five-dimensional Minkowski spacetime
$(\Rnum^5,\flat{ab}) = (\Rnum^4,\fflat{ab})\times(\Rnum,\y{a}{}\y{b}{})$
where $\fflat{ab}$ is the four-dimensional Minkowski metric
and $\y{a}{}$ is a spacelike unit vector orthogonal to $\fflat{ab}$. 
Note we have the 4+1 decompositions
\EQ
\label{decomp}
\flat{ab}= \fflat{ab}+ \y{a}{}\y{b}{}, \quad
\vol{abcde} = 5\fvol{[abcd}\y{e]}{}
\doneEQ
where $\fvol{abcd}$ is the four-dimensional volume form. 
(Throughout, a bar will denote a tensor or field variable on $\Rnum^4$.)
Now we decompose the \grav/ field variables
$\newsymmh{ab}{}{A}$ in the field equations \eqref{5dfieldeq}
into the 4+1 form 
\EQ
\label{symmhdecomp}
\newsymmh{ab}{}{A} = 
\fsymmh{ab}{}{A} + \fA{(a}{A}\y{b)}{} + \fphi{A}\y{a}{}\y{b}{}
\doneEQ
with the fields $\fsymmh{ab}{}{A}$, $\fA{a}{A}$, $\fphi{A}$
taken to have no dependence on the $\y{a}{}$ coordinate:
\EQ
\label{derdecomp}
\y{}{c}\der{c} \fsymmh{ab}{}{A} 
= \y{}{c}\der{c} \fA{a}{A}
= \y{}{c}\der{c} \fphi{A}
=0 . 
\doneEQ
The components of the five-dimensional field equations \eqref{5dfieldeq}
under the decompositions \eqsref{decomp}{derdecomp}
yield four-dimensional field equations consisting of 
$\fE{}{}{A} = \newE{ab}{}{A} \y{}{a}\y{}{b} =0$
for the spin-zero fields $\fphi{A}$, 
and $\fE{a}{}{A} = \newE{ab}{}{A} \y{}{b} - \fE{}{}{A} \y{a}{} =0$
for the spin-one fields $\fA{a}{A}$,
in addition to 
$\fE{ab}{}{A} = 
\newE{ab}{}{A} -2 \fE{(a}{}{A}\y{b)}{} - \fE{}{}{A} \y{a}{}\y{b}{} =0$
for the \grav/ fields $\fsymmh{ab}{}{A}$. 
There is a corresponding decomposition of 
the five-dimensional gauge symmetries \eqref{5dgaugesymm}
in terms of the parameters
$\fvect{}{}{A} = \vect{a}{}{A} \y{}{a}$
and $\fvect{a}{}{A} = \vect{a}{}{A} -\fvect{}{}{A} \y{a}{}$,
which also are taken to have no dependence on the $\y{a}{}$ coordinate, 
\EQ
\y{}{c}\der{c} \fvect{}{}{A}
= \y{}{c}\der{c} \fvect{a}{}{A}
=0 . 
\doneEQ
The resulting four-dimensional gauge theory is a nonlinear deformation of
the combined linear theory of scalar fields 
\EQ
\fEtop{1}{}{A} = -\tfrac{1}{2}\finvflat{cd} \fder{c}\fder{d} \fphi{A} ,\quad
\var{0}{\fvect{}{}{}} \fphi{A} =0 , 
\doneEQ
and Maxwell gauge fields
\EQ
\fEtop{1}{a}{A} = \finvflat{cd} \fder{c}\fder{[d} \fA{a]}{A} ,\quad
\var{0}{\fvect{}{}{}} \fA{a}{A} =\fder{a}\fvect{}{}{A} , 
\doneEQ
and linearized graviton fields
\EQ
\fEtop{1}{ab}{A} 
= -2\finvflat{cd} \fder{[c|}\fder{[d} \fsymmh{b]|a]}{}{A} ,\quad
\var{0}{\fvect{}{}{}} \fsymmh{ab}{}{A} =\fder{(a}\fvect{b)}{}{A} . 
\doneEQ

A Lagrangian formulation is readily obtained by
decomposing the five-dimensional linearized frames
$\h{ab}{}{A} =\invsodder{b}{\mu}\h{a\mu}{}{A}$
and linearized spin-connections 
$\wtop{1}{abc}{}{A} = 
\invsodder{b}{\mu}\invsodder{c}{\nu}\wtop{1}{a\mu\nu}{}{A}$
into the 4+1 form 
\EQs
&& 
\h{ab}{}{A} = 
\fh{ab}{}{A} +\fA{a}{A}\y{b}{} + \fphi{A}\y{a}{}\y{b}{} , 
\label{hdecomp}
\\
&&
\wtop{1}{abc}{}{A} = 
\fw{abc}{}{A} -\fF{bc}{}{A}\y{a}{} +2\fF{a[b}{}{A}\y{c]}{}
-2 \fH{[b}{}{A}\y{c]}{} \y{a}{} , 
\label{wdecomp}
\doneEQs
where 
\EQ
\fw{abc}{}{A} = 3\fder{[a}\fh{bc]}{}{A} -2\fder{[b}\fh{c]a}{}{A} 
\doneEQ
represents the four-dimensional linearized spin-connection,
and 
\EQ
\fF{ab}{}{A} = \fder{[a}\fA{b]}{A} ,\quad
\fH{a}{}{A} = \fder{a}\fphi{A} 
\doneEQ
represent the four-dimensional 
spin-one field strength and spin-zero field strength. 
The linearized spin-connection and field strengths here 
have the role of auxiliary fields 
(analogous to a ``1.5'' formalism in supergravity \cite{supergravity}). 
Then, through the decompositions \eqref{decomp}, \eqrefs{hdecomp}{wdecomp},
the five-dimensional Lagrangian \eqref{5dL} reduces to the form 
$\bar L=\fLtop{2}+\fLtop{3}$
where
\EQs
\fLtop{2} = &&
-\tfrac{1}{2}( 
\fh{a}{a}{A} ( \fder{b}\fw{c}{bc}{B} - \fder{b}\fH{}{b}{B} )
+\fh{b}{a}{A}( 2\fder{[c}\fw{a]}{bc}{B}+ \fder{a}\fH{}{b}{B} )
\nonumber\\&&
-\fw{[a}{ca}{A}\fw{b]c}{b}{B} 
-\fA{b}{A} \fder{c}\fF{}{bc}{B} +\tfrac{1}{2} \fF{ab}{}{A} \fF{}{ab}{B}
) \id{AB}{} 
\doneEQs
and 
\EQs
\fLtop{3} = &&
-\tfrac{1}{2}( 
( \fw{a}{pq}{A} \fw{bpq}{}{B} -2 \fF{a}{p}{A} \fF{bp}{}{B} )\fw{bcd}{}{C} 
-( 4\fw{ab}{p}{A} \fw{cd}{q}{B} +2\fw{a}{pq}{A} \fw{bcd}{}{B} ) \fF{pq}{}{C} 
\nonumber\\&&
-4 \fH{}{p}{A} \fF{ap}{}{B} \fw{bcd}{}{C} 
-4 \fH{(p}{}{A} \fF{a)b}{}{B} \fw{cd}{p}{C} 
)\finvvol{abcd} \c{}{ABC} . 
\doneEQs
This four-dimensional Lagrangian $\bar L=\fLtop{2}+\fLtop{3}$
is invariant to within a total divergence
under the gauge symmetries
\EQs
&&
\delta_{\fvect{}{}{}}\fh{ab}{}{A}= 
\fder{a}\fvect{b}{}{A} ,\quad
\delta_{\fspin{}{}{}}\fh{ab}{}{A}= 
\fspin{ab}{}{A} + 4\c{A}{BC} \fvol{bcpq} \fF{a}{c}{B} \fspin{}{pq}{C} , 
\\
&&
\delta_{\fvect{}{}{}}\fA{a}{A}= 
\fder{a}\fvect{}{}{A} ,\quad
\delta_{\fspin{}{}{}}\fA{a}{A}= 
\c{A}{BC}( \fvol{bcpq} \fw{a}{bc}{B} +2 \fvol{abpq} \fH{}{b}{B} ) 
\fspin{}{pq}{C} , 
\\
&&
\delta_{\fvect{}{}{}}\fphi{A}= 0,\quad
\delta_{\fspin{}{}{}}\fphi{A}= 
-\c{A}{BC} \fvol{cdpq} \fF{}{cd}{B} \fspin{}{pq}{C} , 
\doneEQs
whose parameters $\fvect{}{}{A}$, $\fvect{a}{}{A}$, 
$\fspin{ab}{}{A}=\fspin{[ab]}{}{A}$
are arbitrary functions of the four-dimensional spacetime coordinates, 
as obtained by decomposing the five-dimensional gauge symmetries \eqref{5dL}
and simplifying various terms 
via the nilpotency \eqref{aaeq} of $\c{A}{BC}$. 

Thus we have obtained 
a four-dimensional parity-violating nonlinear gauge theory of 
a massless coupled set of 
\grav/ fields, spin-one fields, and spin-zero fields. 
This theory can be generalized to include 
an algebra-valued gravitational coupling, 
where the internal algebras 
$(\c{A}{BC},\Rnum^N)$ and $(\a{A}{BC},\Rnum^N)$ 
underlying the respective parity-violation coupling 
and gravity coupling in the theory 
satisfy the necessary conditions stated in Theorem~2. 
A simple example for these algebras consists of 
taking the vector space $(\Rnum^6,\id{AB}{}=diag(+1,+1,+1,-1,-1,-1))$
with $\c{}{ABC}$ being the skew product of 
three mutually orthogonal null vectors 
$u\downindex{A},v\downindex{B},w\downindex{C}$,
and $\a{}{ABC}$ being the sum of any symmetric products of these same vectors.

\end{document}